\documentclass[fleqn,usenatbib]{mnras}

\usepackage{newtxtext,newtxmath}

\usepackage[T1]{fontenc}

\DeclareRobustCommand{\VAN}[3]{#2}
\let\VANthebibliography\thebibliography
\def\thebibliography{\DeclareRobustCommand{\VAN}[3]{##3}\VANthebibliography}

\usepackage{graphicx}	
\usepackage{amsmath}	
\usepackage{svg}
\usepackage{tabularx}
\usepackage{calc}
\usepackage{float}
\usepackage{xcolor}
\usepackage{ulem}

\definecolor{red}{rgb}{1.0,0,0}

\title[Habitability of FFP's H$_2$ Exomoons]{Habitability of Tidally Heated H$_2$-Dominated Exomoons around Free-Floating Planets}

\author[Dahlb\"udding \textit{et al.}]{
David Dahlb\"udding$^{1,2,3}$\thanks{E-Mail: ddahlb@mpe.mpg.de},
Tommaso Grassi$^{2,3}$\thanks{Corresponding Author E-Mail: tgrassi@mpe.mpg.de},
Karan Molaverdikhani$^{1,3}$,
Giulia Roccetti$^{4,5,6}$,\newauthor
Barbara Ercolano$^{1,2,3}$,
Dieter Braun$^{7}$
and Paola Caselli$^{2,3}$
\\
$^{1}$Universitäts-Sternwarte, Fakultät für Physik, Ludwig-Maximilians-Universität München, Scheinerstr. 1, 81679 München, Germany; \\
$^{2}$Centre for Astrochemical Studies, Max-Planck-Institut für extraterrestrische Physik, Gießenbachstr. 1, 85749 Garching, Germany; \\
$^{3}$Exzellenzcluster ‘Origins’, Boltzmannstr. 2, 85748 Garching, Germany; \\
$^{4}$European Space Agency (ESA), European Space Astronomy Centre (ESAC), Camino Bajo del Castillo s/n, Villanueva de la Cañada, E-28692 Madrid, Spain;\\
$^{5}$European Southern Observatory, Karl-Schwarzschild-Straße 2, 85748 Garching, Germany;\\
$^{6}$Meteorologisches Institut, Ludwig-Maximilians-Universität München, Theresienstr. 37, 80333 München, Germany and\\
$^{7}$Department of Physics, Center for Nanoscience Ludwig-Maximilians-Universität München, Geschwister-
Scholl Platz 1, 80539 München, Germany;
}

\date{Accepted 2026 February 01. Received 2026 January 26; in original form 2025 October 27}

\pubyear{2026}

\begin{document}
\label{firstpage}
\pagerange{\pageref{firstpage}--\pageref{lastpage}}
\maketitle

\begin{abstract}
Exomoons around free-floating planets (FFPs) can survive their host planet's ejection. Such ejections can increase their orbital eccentricity, providing significant tidal heating in the absence of any stellar energy source. Previous studies suggested that liquid water could exist on such moons under thick CO$_2$-dominated atmospheres, but these models faced challenges with CO$_2$ condensation and atmospheric collapse, particularly in the high-pressure regimes that favoured long-term habitability. To address this, we employ a self-consistent model, including radiative transfer and equilibrium chemistry with condensation, to simulate a more stable hydrogen-dominated atmosphere for a range of initial chemical compositions, including C, O, and N. We find that such atmospheres can effectively trap heat via collision-induced absorption of H$_2$, maintaining surface temperatures suitable for liquid water for time-scales of up to 4.3 Gyr, depending on the surface pressure, while not prone to condensation-induced collapse. Wet-dry cycling caused by the strong tides together with the alkalinity of dissolved NH$_3$ could create favourable conditions for RNA polymerisation and thus support the emergence of life.
\end{abstract}




\section{Introduction}

The latest estimates from microlensing surveys put the population of free-floating planets (FFPs) at about 21 per star \citep{Sumi_2023}. Although this likely overstates their abundance at low masses due to a simple power-law fit, more recent theoretical models predict around 2 ejected FFPs per star for planetary masses above 0.01 M$_\oplus$, with about half of them above 0.33 M$_\oplus$ \citep{Coleman_2025}. Even our Solar System probably contributed at least one ice giant to this FFP population \citep{Nesvorný_2012}.

Future observations will provide us with more data to further constrain the mass function of FFPs. Furthermore, these observations could even detect moons around FFPs by observing transits with JWST \citep{Limbach_2021} or Roman \citep{Limbach_2023, Soares-Furtado_2024} or microlensing lightcurves with Euclid \citep{Bachelet_2022}.

The search for exomoons within conventional stellar systems continues \citep{Teachey_2018, Kipping_2022a, Kipping_2022b} with no confirmed detection to date \citep{Kreidberg_2019, Yahalomi_2023, Heller_2024}. Thus, FFPs might offer an alternative pathway for the first discovery of an exomoon \citep{Limbach_2021}.

Although FFPs could already have the chance of being habitable if a thick atmosphere is able to retain the heat from their formation as well as from radioactive decay \citep{Stevenson_1999}, their moons could have even more favourable conditions for life.

Within our Solar System, the icy moons of both Jupiter and Saturn are prime candidates for harbouring life in oceans below their thick ice crusts \citep{Nimmo_2016}. Europa and Enceladus are especially interesting targets, as their subsurface oceans are in direct contact with the rocky core, possibly providing bio-essential elements \citep{Cockell_2016, Cockell_2024}. These vast bodies of liquid water are made possible not only through radiogenic but also through tidal heating \citep{Nimmo_2016}. Io, the innermost of the Galilean moons, showcases the significance of such tidal forces, as they make it the most volcanically active body in the solar system \citep[e.g.][]{deKleer_2024}.

Although Io is relatively dry and hence unlikely to be habitable \citep{Bierson_2023}, exomoons with a higher water content and a substantial atmosphere could in some cases be more habitable than their planetary counterparts: moons receive less stellar flux due to passing through the shadow provided by their host planet, making them less prone to runaway greenhouse if they are closer to a star than the "classical" habitable zone, as defined in \citet{Kasting_1993}, would allow \citep{Heller_2012}. If, on the contrary, planets are located beyond the habitable zone, the extra energy provided by tidal heating could be crucial to support habitable surface conditions \citep{Scharf_2006, Heller_2013, Dencs_2025}.

The extreme case of this would be a moon around an FFP, without any stellar energy source. \citet{Roccetti_2023} confirmed in line with previous work \citep{Hong_2018, Rabago_2019} that such moons can survive ejections of their host planet from their respective stellar system. They showed that close encounters before the final ejection even increase the ellipticity of the moon's orbit, boosting tidal heating over millions to billions of years, depending on the moon's and FFP's properties.

Building on this, \citet{Avila_2021} and \citet{Roccetti_2023} modelled the potential atmospheres of such rocky moons, arguing that a CO$_2$-dominated atmosphere would represent the optimal case for liquid water at the surface, as it is less likely to escape than species with a lower molecular weight such as H$_2$ and He \citep{Catling_2009}. At the same time, its high opacity provides more of a greenhouse effect than, e.g., N$_2$ \citep{Badescu_2010}, and it also has a more plausible formation history than, e.g., a CH$_4$-dominated atmosphere \citep{Lammer_2018}. Both previous works demonstrated with a simplified radiative and convective temperature profile, employing the Rosseland mean opacities of CO$_2$ from \citet{Badescu_2010}, that surface liquid water would be possible, while \citet{Roccetti_2023} additionally estimated the maximum time-scales of such habitable conditions to reach from a few million to more than a billion years, depending on the surface pressure, which varied from 0.1 to 100 bars.

However, the temperature profiles frequently fall below the condensation curve of CO$_2$, causing atmospheric collapse, particularly in the high surface pressure scenarios that otherwise predicted the longest time-scales for habitable conditions. In this paper, we want to address this problem: First, we ensure accurate temperature profiles by employing the radiative transfer code \textsc{HELIOS} \citep{Malik_2017, Malik_2019a, Malik_2019b, Whittaker_2022} coupled with the equilibrium condensation chemistry code \textsc{GGchem} \citep{Woitke_2018}. We then model a more stable hydrogen-dominated atmosphere with varying abundances of carbon and oxygen. Such incondensible, primordial atmospheres have already been shown to remain habitable beyond the "classical" habitable zone through the potent collision-induced absorption (CIA) of infrared light provided by H$_2$ \citep{Pierrehumbert_2011, MolLous2022}.

Beyond their influence on the surface temperatures of a planet or moon, atmospheres could also aid the emergence of life in another way. To understand how life emerged and how atmospheres could be instrumental to it, we must consider what drives life fundamentally: two of the most important criteria for life include nonequilibrium processes and Darwinian evolution \citep[e.g.][]{Irwin_2020}. If life succumbs to entropy, it dies. Today, a sophisticated biophysical and biochemical machinery maintains this nonequilibrium as long as possible, but during life's first steps, this was not yet available. Hence, some naturally occurring nonequilibrium accumulation process \citep{Braun_2002, Mast_2010, Morasch_2019} should localise and select molecules that can grow in length and undergo exponential selection \citep{Mast_2010, Kreysing_2015, Ianeselli_2022}, setting life up for its core functions of replication and evolution \citep{Kudella_2021}. Of the many possible options currently under investigation, wet-dry cycles, driven by daily or seasonal changes on early Earth, could offer a simple solution as they support early polymerisation and replication of RNA \citep[e.g.][]{Dass_2023, Caimi_2025}. The atmosphere influences this cycle in two ways: First, the pressure, temperature, and chemical composition that it provides drive the evaporation. In addition, its chemical components can dissolve in the wet phase, thus exerting a direct influence on any chemical reactions by, e.g., changing the pH. We want to provide a more in-depth explanation on why wet-dry cycles combined with the right chemical reactions could provide a favourable environment for the very first steps of life. Thereafter, we will consider how this influences our assessment of the habitability of exomoons around FFPs with the specific atmospheric compositions considered in this paper.

In the next section, we will describe our model and assumptions in more detail. The following section will then describe the results alongside their limitations and biochemical implications, all of which are summarised and discussed in the final part of the paper.

\section{Methods}

\subsection{FFP-Exomoon System}
\label{sec:systemSetup}

To be consistent with the previous paper of \citet{Roccetti_2023}, we assumed an Earth-sized planet orbiting a Jupiter-like FFP. While they also considered a more realistic mass distribution of the moons from the population synthesis model of \citet{Cilibrasi_2021}, the less massive moons have a lower chance of entering the habitable zone because the tidal heating energy flux depends on the moon's mass, and at the same time they are unlikely to retain a dense atmosphere due to their lower gravity. Hence, an Earth-like exomoon presents a plausible best-case scenario.

Furthermore, this consistency allowed us to subsequently calculate the time spent in the habitable zone using the distribution of semi-major axes and eccentricities from \citet{Roccetti_2023}. These orbits, originally from the population synthesis study of \citet{Cilibrasi_2021}, were subjected to the ejection of their host planet, ejecting some of the moons in turn while only increasing the eccentricities of others. For the case of a last close encounter with another Jupiter-like planet with an impact parameter of 15.36 $R_\mathrm{J}$ ("Sim1" in \citet{Roccetti_2023}), a total 6945 of the initial 26293 Earth-mass moons survived. Afterwards, the orbits were evolved according to the time-dependent tidal heating differential equations of \citet{Bolmont_2011}, ultimately leading to the circularisation of all individual orbits in the absence of any interaction between moons.

The resulting effective temperature of the moon corresponds to the internal temperature in our models of its atmosphere, described in the next section. 

\begin{table*}
    \caption{The sources for all considered opacities as included in HELIOS or downloaded from DACE \citep{Grimm_2021}}
    \label{tab:opacities}
    \begin{tabularx}{0.7\textwidth}{llX}
        \hline
        Species & Opacity Source & Reference \\
        \hline
        \hline
        CH$_4$ & YT34to10 & \citet{Yurchenko_2014_CH4a, Yurchenko_2017_CH4b}\\
        H$_2$O & POKAZATEL & \citet{Polyansky_2018_H2O}\\
        CO     & Li2015 & \citet{Li_2015_COa, Somogyi_2021_COb}\\
        CO$_2$ & UCL-4000 & \citet{Yurchenko_2020_CO2}\\
        C$_2$H$_2$ & aCeTY & \citet{Chubb_2020_C2H2}\\
        NH$_3$ & CoYuTe & \citet{AlDerzi_2015_NH3a, Coles_2019_NH3b}\\
        HCN    & Harris & \citet{Harris_2006_HCNa, Barber_2013_HCNb}\\
        CIA H$_2$-H$_2$ $^\ast$  & HITRAN & \citet{Abel_2011_CIA_H2-H2a, Fletcher_2018_CIA_H2-H2b}\\
        \hline
        \multicolumn{2}{l}{$^\ast$ F18 for $\lambda \gtrsim 18.5 \mu$m and $T \leq 400$K}
    \end{tabularx}
\end{table*}

\subsection{Self-Consistent Model}
\label{sec:model}

In a first step towards a comprehensive self-consistent atmospheric model, we employed the radiative transfer code \textsc{HELIOS} to accurately model the temperature profile and ensure the stability of the atmosphere. A key advantage of this approach is the possible addition of further molecular opacities. The Rosseland mean opacities of individual molecules used in previous models cannot simply be summed up for a mixture of gases, as one would have to recalculate the integral over frequency \citep[Eq. (17) of][]{Badescu_2010}. Another advantage of HELIOS is the possibility of handling surface layers of rocky planetary bodies.

Additionally, \textsc{HELIOS} can be coupled to a code that predicts the abundances of the various chemical species given the specific physical conditions. In our case, this is the equilibrium condensation chemistry code \textsc{GGchem}, which offers numerical stability down to temperatures as low as 100 K - a key capability in our starlight-deprived worlds.

Both codes operate under a set of assumptions: \textsc{HELIOS} calculates the atmospheric structure assuming constant gravity, which can lead to inaccurate results, especially for thick atmospheres ($P_\mathrm{surf} = 100$ bar) on low-gravity moons (e.g., Io). In this regime, the atmosphere's vertical extent is substantial compared to the moon's radius, causing gravity to vary considerably with altitude. It also does not consider the influence of condensation on the temperature profile, which could otherwise follow a moist adiabat (although this feature was recently added \citet{Tsai_2024}\footnote{\url{https://github.com/exoclime/HELIOS/tree/development}}). \textsc{GGchem} calculates the equilibrium chemistry in each layer independently, i.e., each atmospheric layer is modelled in a one-zone fashion. No atoms or molecules are redistributed to different layers, neither through diffusion nor through rainout. Because the surface temperature mainly depends on a narrow layer around the radiative-convective boundary, vertical mixing should not significantly affect our results.

The iterative coupling process begins with an initial isothermal temperature profile (500 K) to compute the corresponding equilibrium molecular abundances using \textsc{GGchem}. This relatively high temperature regime ensures that no significant condensation occurs at first, which could decrease the optical thickness of the atmosphere, leading to a runaway cooling process. While these initial conditions are foremost motivated by stability, they can also be related to the hot conditions after moon formation. The first iteration of \textsc{HELIOS} runs for a maximum of 1000 iterations, even if convergence is not reached, to obtain a better first approximation than the initial constant 500 K. The subsequent iterations between \textsc{GGchem} and \textsc{HELIOS} are run a maximum of 10 times each, which is usually sufficient for convergence, apart from the upper layers above 10$^{-5}$ bar, which do not affect the surface temperature. \textsc{HELIOS} has the option to turn on "coupling speed up", which simply averages the last two solutions. While this prevents oscillations, we found that it is best to enable it only after a few steps, as it can otherwise also hinder convergence. We also set the convergence criterion, defined as the relative temperature change between subsequent HELIOS runs, to 10$^{-3}$. Other parameter choices as well as the full coupling code can be found in the corresponding repository \footnote{\url{https://github.com/DavidDahlbudding/chelio}}.

A list of the used opacities and their sources are listed in Table \ref{tab:opacities}. They span a temperature range of 50 - 6000 K and pressures from 10$^{-6}$ to 10$^{3}$ bars with a resolution of $R := \lambda / \Delta\lambda = 50$, saved in the correlated-k format. Because the total opacity changes as the molecular abundances vary between iterations, we calculated it on the fly using random overlap for mixing. This assumes that the absorption lines of the mixed molecules are uncorrelated, which should yield more accurate results than assuming the opposite \citep{Amundsen_2017}.

As a final check, we calculated the stability of each atmosphere with respect to Jeans escape by estimating the time-scale of the whole atmospheric mass being lost with a constant escape flux. This Jeans escape flux is calculated according to Eqs. (4.50) and (4.54) in \citet{Seager_2010}. If the atmosphere had to be extended to find the radius of the exobase, we assumed a constant temperature and gravity, consistent with \textsc{HELIOS}.

\subsection{Parameter Space Exploration}

\begin{table}
    \caption{The ranges or steps used for parameter exploration}
    \label{tab:parameters}
    \begin{tabularx}{\columnwidth}{XX}\hline
        \hline
        Parameter & Range/Steps\\
        \hline
        \hline
        $P_\mathrm{surf}$ [bar] & [1, 10, 100] \\
        $T_\mathrm{int}$ [K] & [50, 250] \\
        $X_{\mathrm{C+O}}$ &  [10$^{-3}$, 10$^{-1}$] \\
        C/O & [0.1, 0.59, 1.0] \\
        g* [$\frac{\mathrm{cm}}{\mathrm{s}^2}$] & [180, \textbf{981}]\\
        $X_\mathrm{N}$* & [\textbf{0}, 10$^{-4}$, 10$^{-2}$] \\
        \hline
        \multicolumn{2}{l}{$^\ast$ additional parameters only varied for one particular case}\\
        \multicolumn{2}{l}{\hspace{\widthof{$^\ast$ }}(default value marked \textbf{bold})}
    \end{tabularx}
\end{table}

The main difference from previous papers is that we investigated hydrogen-dominated atmospheres. As briefly mentioned in the introduction, this has the key advantage of making the atmosphere more stable because H$_2$ is not readily condensable. Additionally, the collision-induced absorption of hydrogen is an effective infrared absorber, and hence can efficiently trap the heat produced by the tides. Although atmospheres with such a low mean molecular weight can easily evaporate under the intense radiation of a young star, ejected FFPs and their moons could be able to retain even the most massive ones \citep{MolLous2022}.

To investigate how different minor species can contribute to or hinder this greenhouse effect, we varied the abundance of carbon and oxygen. First we set their combined volume mixing ratio $X_\mathrm{C+O} := X_\mathrm{C} + X_\mathrm{O}$ to between 10$^{-3}$, roughly the solar value \citep[$= 7.8 \times 10^{-4}$, see][]{Asplund_2020}, and 10$^{-1}$, where other CIA opacities could become important. We then adjusted the abundances to match a specific C/O ratio, varied between 0.1 and 1.0 with the solar value at 0.59 \citep{Asplund_2020}.

We also investigated how other parameters influence the surface temperature. We varied the surface gravity to examine the trade-off between a higher column density of the atmosphere at constant surface pressure and a lower heating rate resulting from a smaller moon. Secondly, to further analyse how diverse chemical species can contribute to the greenhouse effect, we introduced nitrogen into the system, the main component of our atmosphere and crucial for the emergence of life \citep{Cockell_2016}. Specifically, we added the rounded solar quantity \citep[$= 6.8 \times 10^{-5}$, see][]{Asplund_2020} of $X_\mathrm{N} = 10^{-4}$ and an enhanced value of $10^{-2}$.

In this work, we define the habitability of a moon by checking that the surface temperature is in the range for water to be liquid at the respective surface pressure \citep{Avila_2021, Roccetti_2023}. This may overestimate the habitable zone, considering that life on Earth can survive only up to $\sim 150^\circ$C, even in high-pressure environments \citep{Merino_2019}.

In principle, we want to achieve these conditions with as low an internal temperature as possible. As both the tidal heating energy flux and the dissipation time-scale have a strong dependence on the semi-major axis of the moon's orbit, we want to keep the tidal heat to a minimum to maximise the time-scale over which it can be maintained \citep[see Eq. (6) and (16) of][]{Roccetti_2023}. Therefore, expanding the habitable zone to temperatures above $100^\circ$C at high pressures does not influence this maximum time-scale.

\section{Results}

\begin{figure}
\begin{center}
    \begin{minipage}{.24\textwidth}
        \includegraphics[width=\textwidth]{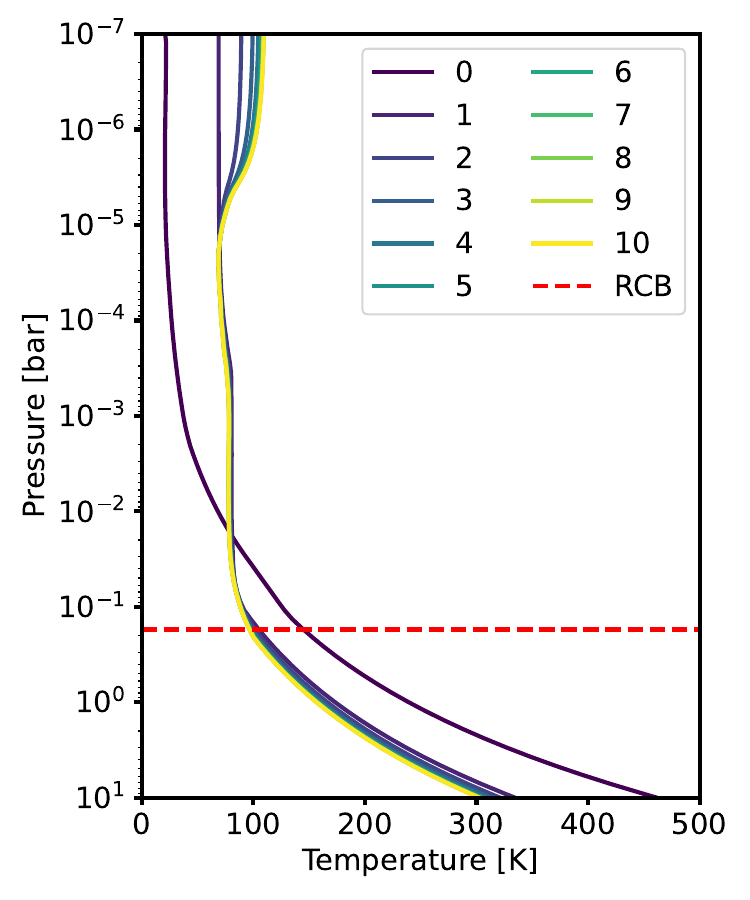}
    \end{minipage}%
    \begin{minipage}{.24\textwidth}
        \includegraphics[width=\textwidth]{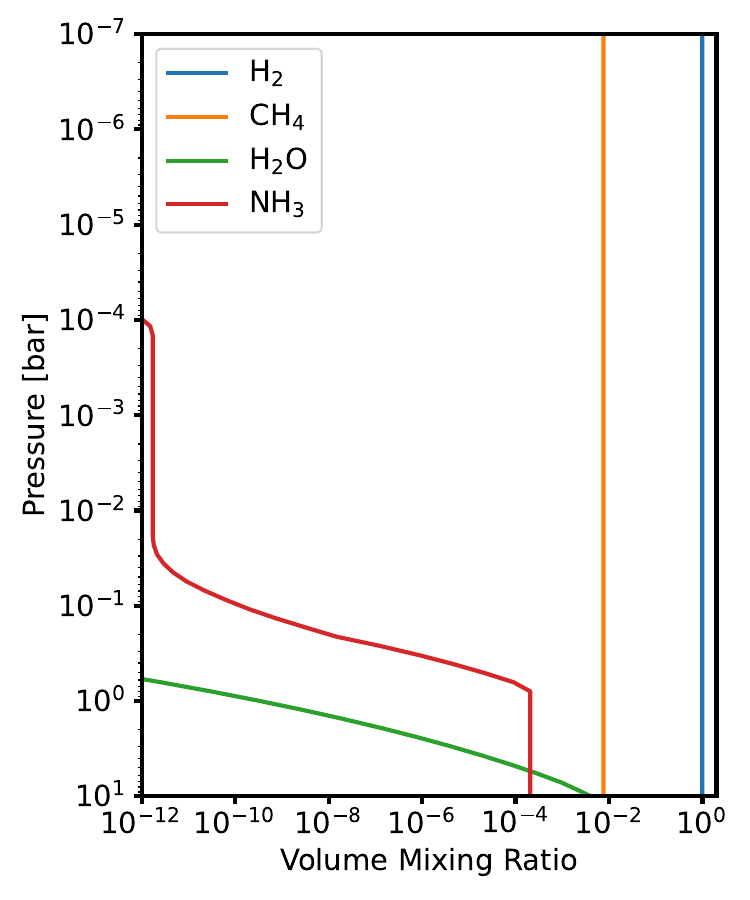}
    \end{minipage}%
    \caption{Example run of the coupled HELIOS-GGchem code with P$_\mathrm{surf}$ = 10 bar, T$_\mathrm{int}$ = 100 K, $X_\mathrm{C+O}$ = 10$^{-2}$, C/O = 0.59 and $X_\mathrm{N}$ = 10$^{-4}$. Left: Evolution of the pressure-temperature profile T(P) with the radiative-convective boundary (RCB) of the last step. Right: Final VMR profiles of all molecules with a maximum VMR > 10$^{-12}$.}
    \label{fig:iteration}
\end{center}
\end{figure}

In Fig. \ref{fig:iteration} we show an example run of the coupled \textsc{HELIOS}-\textsc{GGchem} code. In the left plot, we can see the temperature profile convergence. The plot on the right shows the volume mixing ratios (VMRs) of all relevant molecules resulting from the final temperature profile.

These VMR profiles are representative for all models: In most cases, the combinations of pressure and temperature we found produce CH$_4$ in the gas phase, determining the constant profile shown in Fig. \ref{fig:iteration}. Although CH$_4$ can condense at the lowest temperatures encountered, its limiting factor is primarily the total abundance of carbon available in the system. The H$_2$O abundance, on the other hand, is limited by its condensation curve and therefore has its maximum abundance near the surface, where the temperature peaks. In the absence of any atomic species beyond H, C, and O or any nonequilibrium chemical processes, these three molecules - hydrogen, methane, and water - are the only ones present in relevant quantities.

\begin{figure*}
    \centering
    \begin{minipage}{.33\textwidth}
        \centering
        \includegraphics[width=\textwidth]{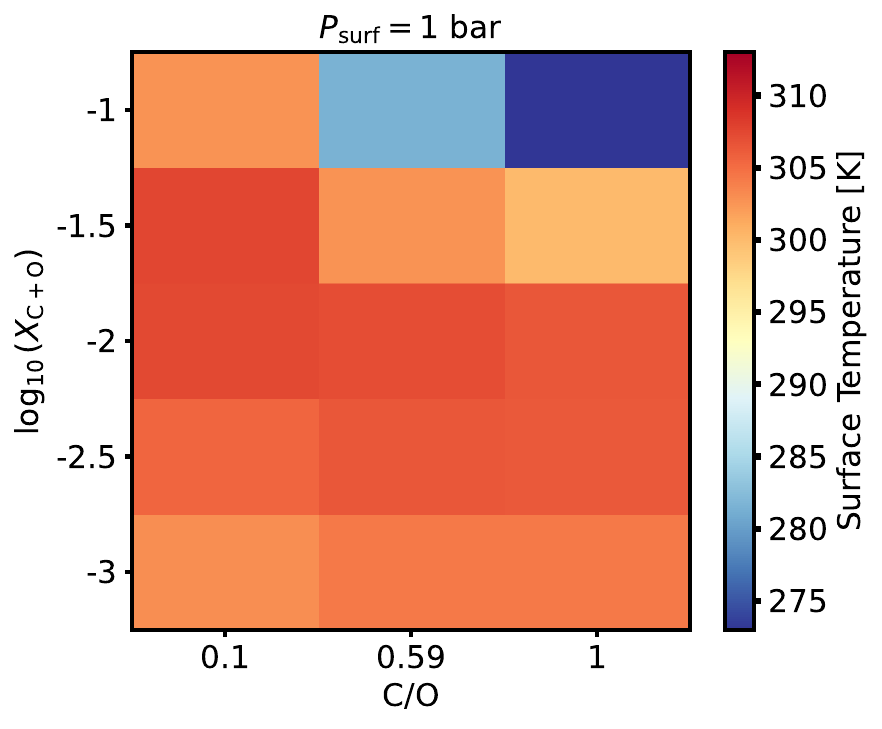}
    \end{minipage}%
    \begin{minipage}{.33\textwidth}
        \centering
        \includegraphics[width=\textwidth]{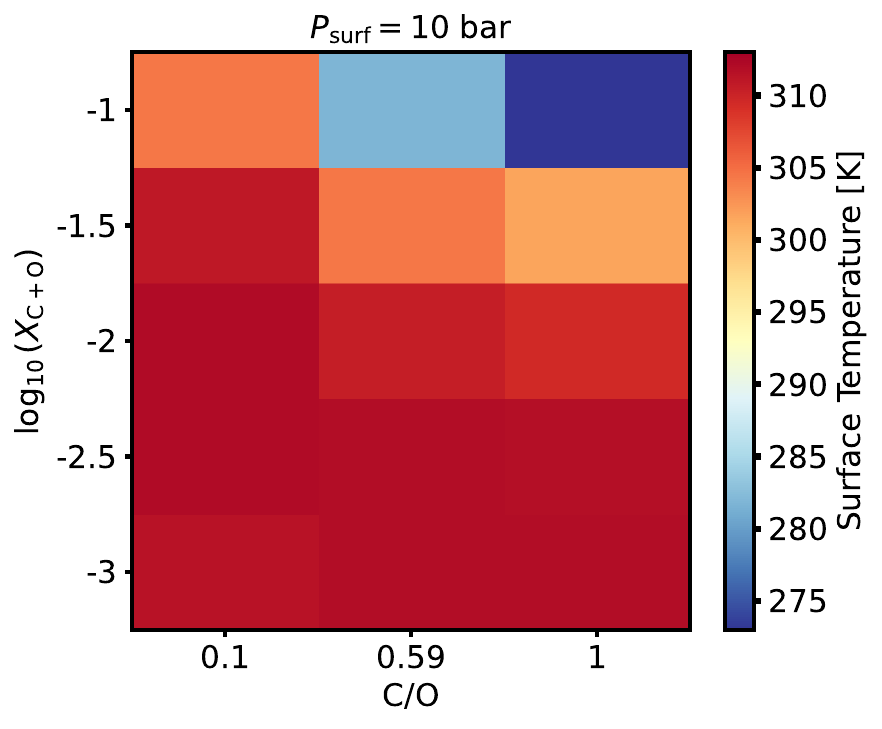}
    \end{minipage}%
    \begin{minipage}{.33\textwidth}
        \centering
        \includegraphics[width=\textwidth]{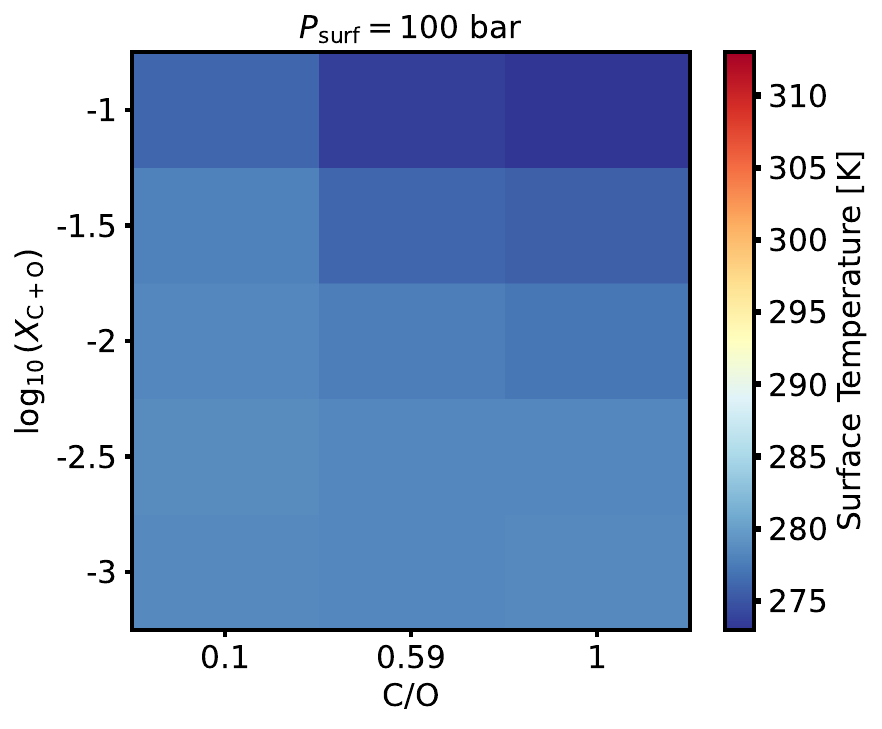}
    \end{minipage}%
    \caption{Surface temperatures T$_\mathrm{surf}$ ($X_\mathrm{C+O}$, C/O) for varying P$_\mathrm{surf}$ (from left to right: 1, 10, and 100 bar). For comparability, the internal temperatures T$_\mathrm{int}$ (185, 104, and 56 K, respectively) were chosen such that the minimum T$_\mathrm{surf}$ for each pressure is at the freezing point of water, 273 K. For this, T$_\mathrm{surf}$(T$_\mathrm{int}$) was linearly interpolated.}
    \label{fig:BOA_temps}
\end{figure*}

Fig. \ref{fig:BOA_temps} shows the surface temperatures depending on the composition of the atmospheres for the three different surface pressures of 1, 10, and 100 bar. The value of each internal temperature - 185, 104, and 56 K, respectively - was chosen such that the lowest interpolated surface temperature is still above freezing.

This lowest surface temperature can typically be found in the high $X_\mathrm{C+O}$ (=10$^{-1}$) and high C/O (=1.0) case, where most carbon and hence most CH$_4$ is present. It mainly influences the surface temperature by decreasing the H$_2$ abundance: Because the absorption coefficient of the CIA scales with the density of H$_2$ squared, we observe surface temperatures decreasing with increasing $X_\mathrm{C+O}$. This also explains the shift of the maximum temperature to lower $X_\mathrm{C+O}$, as the surface pressure increases and the H$_2$-H$_2$ CIA gains relative importance. That is also why temperature differences are less pronounced in the high-pressure case.

Water condenses sharply with increasing height in the atmosphere (cf. Fig. \ref{fig:iteration}). Especially in high-pressure cases, this depletes water, rendering it optically insignificant upon reaching the radiative-convective boundary. Only for the lowest surface pressure of 1 bar, H$_2$O remains optically relevant up to this key boundary, which is why we see a slight decrease of surface temperatures towards low $X_\mathrm{C+O}$.

\begin{figure}
    \includegraphics[width=.45\textwidth]{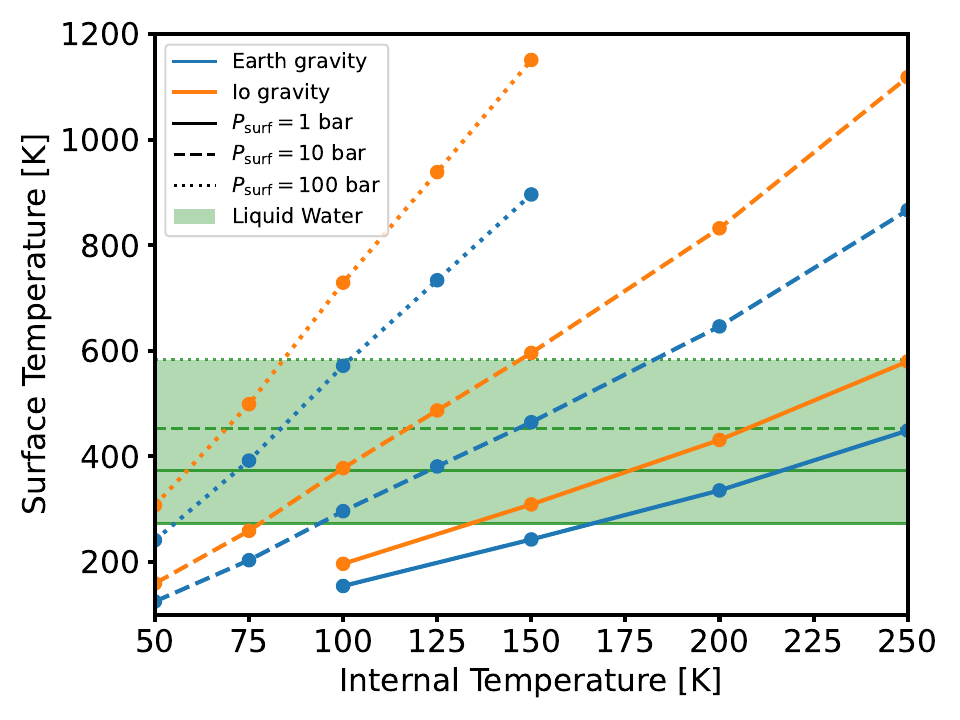}
    \caption{The surface temperature T$_\mathrm{surf}$ increasing with the tidally provided internal temperature T$_\mathrm{int}$ for Earth- vs. Io-like gravity for $X_\mathrm{C+O}$ = 10$^{-2}$ and C/O=0.59. In a lower-gravity environment, more atmospheric mass is required to maintain a constant surface pressure P$_\mathrm{surf}$, resulting in an optically thicker and hence hotter atmosphere.}
    \label{fig:ComparisonPlanet}
\end{figure}

Decreasing the surface gravity necessitates a larger atmospheric column mass to produce the same surface pressure. This also increases the optical depth of the atmosphere and ultimately contributes to higher surface temperatures, as we see in Fig. \ref{fig:ComparisonPlanet}. The habitable zone for the case of an Io-like gravity is already reached at 134, 78, or 46 K of internal temperature for the surface pressures of 1, 10, and 100 bar, respectively, compared to 167, 94, or 55 K for an Earth-like moon, i.e., an 18\% decrease on average.

However, a smaller moon also produces less tidal heat: Using Equations (14) and (15) from \citet{Roccetti_2023}, we can roughly estimate the scale of this process with T$_\mathrm{int} \propto R_\mathrm{m}^\frac{3}{4}$, where $R_\mathrm{m}$ is the radius of the moon. Given that the radius of Io is approximately 28.6\% that of Earth, this scaling implies a reduction of roughly 61\% in the tidally produced internal temperature for an Io-sized moon compared to an Earth-sized moon on the same orbit.

\begin{figure}
    \includegraphics[width=.45\textwidth]{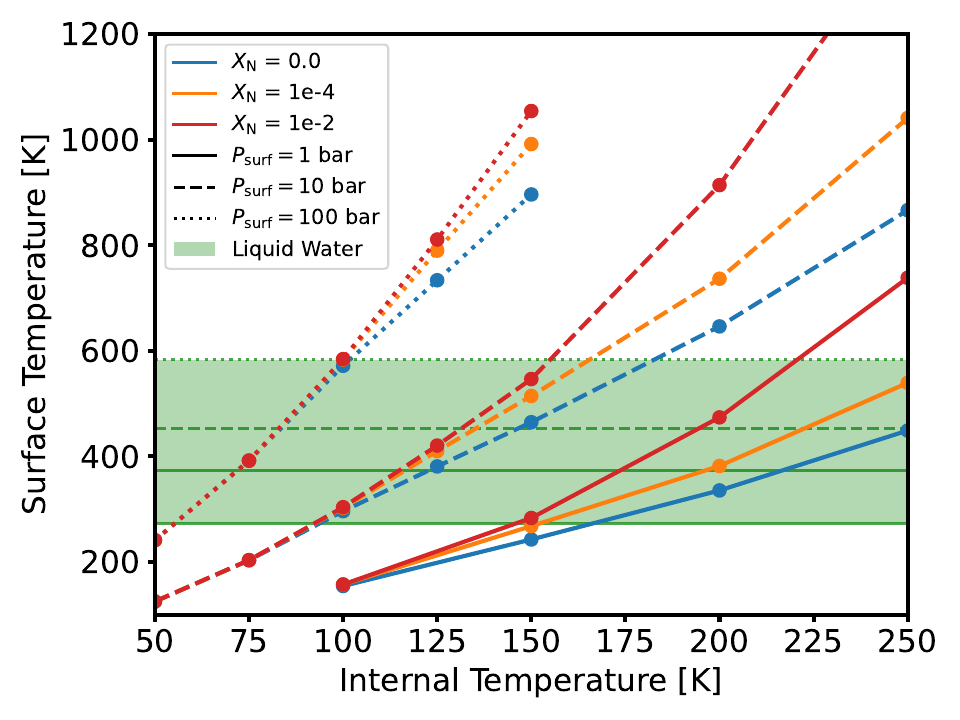}
    \caption{The surface temperature T$_\mathrm{surf}$ increasing with the tidally provided internal temperature T$_\mathrm{int}$ for varying abundances of N for $X_\mathrm{C+O}$ = 10$^{-2}$ and C/O=0.59. The produced NH$_3$ only increases temperatures if it can reach beyond the optically thick convective part of the atmosphere before condensing, thereby shifting the radiative-convective boundary (cf. RCB in Fig. \ref{fig:iteration}).}
    \label{fig:ComparisonN}
\end{figure}

Not only gravity can increase the surface temperature. Also other chemical species, depending on their abundance and absorption spectra, can contribute to the greenhouse effect, as shown in Fig. \ref{fig:ComparisonN}. Adding nitrogen to this hydrogen-dominated atmosphere produces mainly ammonia, which seems to increase the surface temperatures only above a certain internal temperature threshold.

This can be explained by the distribution of NH$_3$ in the atmosphere. Although it is also constrained by condensation similar to H$_2$O, it remains abundant at higher layers than H$_2$O does, as shown in Fig. \ref{fig:iteration}. Still, this increases the surface temperature only if ammonia is present above the radiative-convective boundary, which necessitates a sufficiently high internal temperature.

\begin{figure}
    \includegraphics[width=.45\textwidth]{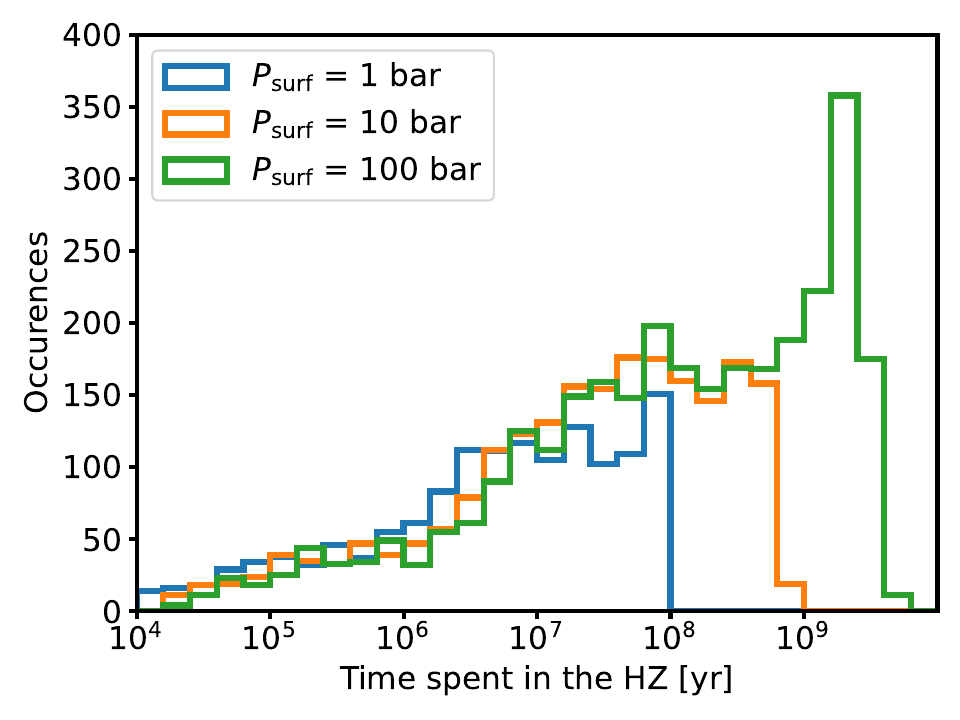}
    \caption{Updated time spent in habitable zone for $X_\mathrm{C+O}$ = 10$^{-2}$ and C/O=0.59 using the semi-major axis and eccentricity distribution and their evolution from \citet{Roccetti_2023} for Earth-mass moons. Of the total sample of 6945 moons [20, 31, 43] \% reach the HZ at any point during their evolution and stay there for a maximum time of [95, 699, 4341] Myr for the surface pressures of [1, 10, 100] bar.}
    \label{fig:Update_TimeInHZ}
\end{figure}

Finally, Fig. \ref{fig:Update_TimeInHZ} shows the updated times spent in the habitable zone by the Earth-mass moons for one of the atmospheric compositions, where $X_\mathrm{C+O}$ = 10$^{-2}$ and C/O = 0.59, using the semi-major axis and eccentricity distribution as well as the tidal evolution model from \citet{Roccetti_2023}. Of the total 6945 moon orbits which survived the ejection of their FFP, 1041 (20.2\%), 2131 (30.7\%) and 2984 (43.0\%) can retain liquid water at some point during their orbital evolution for the respective surface pressures of 1, 10, and 100 bar. The maximum times reached are 95, 699, and 4341 Gyr , respectively. Although on similar scales, these times are significantly longer than the 52, 276, and 1590 Myr found in \citet{Roccetti_2023}.

All atmospheres were also assessed to be stable with respect to Jeans escape, as the respective escape time-scales were found to exceed the lifetime of the universe by far.

\subsection{Limitations}

As described in the Methods section, several approximations and assumptions go into \textsc{HELIOS} and \textsc{GGchem}. This includes that only dry, but not moist, adiabats are considered when calculating the temperature profile, which could further increase surface temperatures by decreasing the temperature-pressure gradient in the relevant layers where, e.g., water or ammonia is condensing.

Another effect of condensation is the formation of clouds. Enough cloud coverage could trap even more heat, and hence, even less tidal heating would be necessary for surface liquid water. Hazes, possibly created by volcanic activity, could have a similar effect, a promising avenue to explore in future work.

More generally, orbital resonances could stabilise elliptical orbits, as can be observed in the case of the Galilean moons \citep[e.g.,][]{Sinclair_1975}. Considering this, the estimates for the habitable lifetime shown in Fig. \ref{fig:Update_TimeInHZ}, based on modelling the evolution of individual moon orbits by \citet{Roccetti_2023}, can therefore be seen as lower limits.

At the same time, a thick surface ocean layer could also enhance tidal dissipation \citep{Auclair-Desrotour_2018}, leading to faster circularisation of the moon's orbit and decreasing the time spent in the habitable zone.

It also remains questionable whether thick atmospheres with low mean molecular weight, such as the hydrogen-dominated ones presented here, can form on low-mass moons \citep{Bierson_2023}. If they can, exomoons around free-floating planets, ejected as early as possible from their stellar system, would have the best chance of retaining them.

\subsection{Biochemical Background and Implications}

As outlined in the introduction, wet-dry cycles could have offered a favourable environment for life to take its first steps on Earth.

Complex organic molecules have been detected on the asteroids Ryugu and Bennu \citep{Oba_2023, Glavin_2025}. This is why current research on these first steps of life focuses not on the synthesis of such molecules but on systematically increasing the length and complexity of, e.g., RNA through physical selection pressures \citep[e.g.,][]{Mast_2010, Kreysing_2015, Ianeselli_2022}. While the focus on RNA may seem Earth-centric, the detection of RNA bases on these asteroids suggests that other planetary bodies may experience similar processes. RNA also plays an important role as part of a complex autocatalytic network in the ribosome, which assembles proteins from amino acids transported by tRNA according to instructions encoded in mRNA and is fundamental to life as we know it. During the emergence of life on Earth, physical environments and selection pressures could have created a similar but simpler network to accumulate and polymerise smaller molecules, which were then selectively replicated through templated ligation \citep{Kudella_2021}.

For these processes to work, the ends of the strands must be continuously activated and deactivated. The simplest form of nucleotide activation is 2',3'-cyclic phosphate. This end group is formed when polymerised RNA, connected at positions 3' and 5', breaks up into pieces due to hydrolysis. Further hydrolysis forms deactivated 2' or 3' linear phosphates, which are easily reactivated by the removal of water, triggering polymerisation \citep{Verlander_1973}. The resulting templated ligation replicates strands with a high enough precision for preservation \citep{Eigen_1971,Calaca_2024}.

This scenario resembles the occurrence of wet-dry cycles: In order to achieve RNA polymerisation and replication by ligation, wet and dry conditions must coexist in one location to prevent the diffusion of molecules. Such settings can involve temperature differences \citep{Braun_2002, Mast_2010, Kreysing_2015} or dry states in the form of gas bubbles \citep{Morasch_2019, Tekin_2022}. These combinations have been shown to replicate and select with the help of polymerase proteins \citep{Ianeselli_2022}, even in the absence of temperature differences \citep{Schwintek_2024}. Natural environments offering these conditions are networks of porous rocks, which have been shown to provide a wide range of molecular diversity \citep{Matreux_2024}.

Such reactions are performed with minimal additional salts and with no or minimal Mg$^{+2}$. These conditions enable RNA to survive long-term in the alkaline pH 9–10 environment necessary for polymerisation and templated ligation. Arguments that 2',3'-cyclic phosphate nucleotides are a good starting point are further supported by evidence that amino acids catalyse RNA polymerisation by a factor of more than 100 \citep{Rout_2025}. Further experiments are being conducted to identify prebiotic phosphorylation agents that can recycle back to 2',3'-cyclic phosphate, and to understand how homochiral RNA is created by polymerisation.

To summarise, the 2',3'-cyclic phosphate is readily activated and deactivated through wet-dry cycles, supporting polymerisation and ligation. They are a natural byproduct of RNA decay and thus could be recycled efficiently. Importantly, these reactions are catalysed by amino acids, suggesting the possible emergence of autocatalytic networks that are fundamental to life.

The main influence of the atmospheric composition is regulating the pH. A potent alkalising agent would be required to raise the pH to 9–10, especially in the presence of acidifying CO$_{2}$. One possibility is the role of NH$_{3}$, which provides an alkaline pH in pure water in the gas phase even at partial pressures as low as 0.001 mbar. Notably, NH$_{3}$ can be stored in a dry state as ammonium ions or ammonium carbamates and released directly into the gas phase.

In our models with the lower (solar) value of $X_\mathrm{N} = 10^{-4}$, ammonia reaches partial pressures of 0.2 to 20 mbar, depending on the overall surface pressure. This condition should be sufficient to create adequate alkaline conditions, even with the limiting factor of a significant amount of nitrogen locked up in N$_{2}$ and the presence of CO$_{2}$ through outgassing. Although our exomoons are not subjected to day and night cycles, the variable strength of the tidal force throughout their orbits, in combination with a shallow ocean, could create the necessary wet-dry cycles with periods in the order of days to months.

Note that such cycles assume the existence of an ocean, shallow enough for the existence of dry land. But, as the internal temperature is only weakly dependent on the interior structure and emissivity, and the atmosphere does not chemically interact with the surface, our model is agnostic to the exact composition of the moon and its surface. Adding such an interior model and surface layer could certainly improve future models, but they are beyond the scope of this paper.

\section{Discussion and Conclusion}

In this paper, we employed a coupled radiative transfer and equilibrium chemistry code to model hydrogen-dominated atmospheres of exomoons around free-floating planets. Previous work \citep{Avila_2021, Roccetti_2023} assessed that conditions suitable for liquid water could exist on such moons given sufficient tidal heat and a thick atmosphere, which was assumed to be CO$_2$-dominated. Because CO$_2$ can be prone to condensation in such low-temperature environments, we now modelled H$_2$-dominated atmospheres instead, while also accounting for the crucial aspect of the stability of any atmosphere present.

As originally proposed by \citet{Stevenson_1999}, thick hydrogen-dominated envelopes can effectively trap any internal heat through their collision-induced absorption (CIA), while guaranteeing stability. We confirm this while exploring a range of C and O abundances:

\begin{itemize}
    \item CH$_4$ is the most abundant minor species, with a constant mixing ratio throughout the atmosphere for habitable conditions at the surface
    \item H$_2$O is usually most abundant near the surface but decreases rapidly with height because of its condensation
    \item CIA by H$_2$-H$_2$ remains the most important absorber even in the presence of these other effective greenhouse gases
\end{itemize}

The last point here is evident from the negative correlation of surface temperatures with both the combined abundance $X_\mathrm{C+O}$ and the C/O ratio: These would both result in a higher methane mixing ratio, but the formation of methane ultimately only reduces the density of H$_2$ and therefore the CIA, overshadowing any effect of increased absorption by CH$_4$, especially for high density, i.e. high pressure atmospheres. Water, on the other hand, remains largely present only in convective layers. There, significantly below the radiative-convective boundary, the atmosphere is optically thick, with or without H$_2$O, and hence it has a limited effect on the surface temperature.

The resulting times spent in the habitable zone for one of the atmospheric configurations peaked at 4.3 Gyr for a surface pressure of 100 bar, which is nearly the current age of Earth.

Lowering the moon's gravity while maintaining surface pressure increased the surface temperature for a given internal heat flux, as expected. However, smaller moons also face challenges, producing less tidal heat for a given orbit and potentially having a greater difficulty retaining thick atmospheres over geological time-scales.

We also added one additional atomic species, nitrogen, with varying abundances. This increased surface temperatures only above certain internal temperature thresholds, that is, if NH$_3$ could reach high enough in the atmosphere, above the convective layers, before being cut off by condensation similar to water.

The abundant ammonia could be crucial for the very first steps of life: selectively increasing the length and complexity of RNA strands and subsequently creating the first autocatalytic networks, possibly enabled by wet-dry cycling. The strong tides experienced by the moons could create similar cycles. During these, the atmosphere would act as a chemical buffer, storing NH$_3$ that can dissolve in water to create the alkaline conditions necessary for polymerisation and templated ligation. More generally, chemically reducing environments, such as the H$_2$-dominated atmospheres modelled here, were often present on early Earth after impacts of large asteroids and were likely essential to the origin of life \citep{Benner_2020, Zahnle_2020}.

The present-day Earth looks much different from the worlds presented here, which, with their thick hydrogen envelopes and possibly deep oceans, resemble a Hycean planet. Although usually in the sub-Neptune range, these worlds are prime candidates for the detection of life \citep{Madhusudhan_2021,  Madhusudhan_2023a, Madhusudhan_2023b}. In their case, any tidal heating could conversely narrow the habitable zone \citep{Livesey_2025}. Our small-scale Hycean worlds could provide relatively better conditions for life. Due to their ($\sim 25\%$) lower gravity, high-pressure ices between a potential liquid water ocean and the rocky core would be less likely, allowing the ocean to receive essential nutrients \citep{Cockell_2024}. Although, as \citet{Madhusudhan_2023a} note, this represents only one possible source of these essential biological elements.

In summary, this work presents a successful application of our coupled radiative transfer and chemistry model to hydrogen-dominated atmospheres, specifically designed to handle the low temperature conditions where CO$_2$ condensation becomes problematic. We confirmed that H$_2$-H$_2$ CIA is key to trapping heat and maintaining stability in such environments, enabling potentially habitable conditions despite the lack of incident stellar radiation.

These potentially habitable moons could be detected through a variety of techniques, auch as transits of its FFP \citep{Limbach_2021} or microlensing \citep{Bachelet_2022}. The direct observation of volcanic hotspots could even confirm the absence of a thick atmosphere \citep{Kleisioti_2024}. To verify and analyze an atmosphere, on the other hand, may not be feasible with any instruments currently in operation.

In future work, we will explore habitable configurations beyond a hydrogen-dominated atmosphere and test whether they are stable and can trap sufficient heat. Increasing the complexity of the model by adding more CIA opacities, clouds, and support for moist adiabats - crucial in the presence of significant condensation - will allow us to better assess the habitability of these unseen worlds.

\section*{Acknowledgements}

The authors would like to thank the anonymous reviewers for their constructive comments and valuable suggestions, which helped improve the quality of this manuscript.
This work was supported by the Deutsche Forschungsgemeinschaft (DFG, German Research Foundation), TRR 392 (CRC 392) "Molecular Evolution in Prebiotic Environments" - 521256690, Research Unit “Transition discs” - 325594231 and the Excellence Cluster ORIGINS funded by Germany’s Excellence Strategy EXC-2094 - 390783311.

\section*{Data Availability}

The code, which couples HELIOS and GGchem, as well as the Jupyter Notebooks used to analyse the data, can be found on GitHub: \url{https://github.com/DavidDahlbudding/chelio}.\\
The produced data can be downloaded from Zenodo:\\ \url{https://doi.org/10.5281/zenodo.15738536}.



\bibliographystyle{mnras}
\bibliography{example} 






\bsp	
\label{lastpage}
\end{document}